\documentclass[prc,twocolumn,twoside,showpacs,nofootinbib,floatfix,superscriptaddress]{revtex4}

\usepackage{graphicx,color,rotating}
\usepackage{amsmath,amssymb,bm}
\usepackage{times,txfonts}
\usepackage{pifont}

\newcommand{\eq}[1]{\begin{equation}#1\end{equation}}
\newcommand{\eqmulti}[1]{\begin{equation}\begin{split}#1\end{split}\end{equation}}

\newcommand{\bra}[1]{\ensuremath{\langle{#1}|\,}}

\newcommand{\ket}[1]{\ensuremath{\,|{#1}\rangle}}

\newcommand{\matrixe}[3]{\ensuremath{\langle{#1}|\,{#2}\,|{#3}\rangle}}

\newcommand{\op}[1]{\ensuremath{#1}}

\renewcommand{\vec}[1]{\ensuremath{\bm{#1}}}

\newcommand{\clebsch}[6]{\ensuremath{\,\mathrm{c}\Big( 
\begin{smallmatrix} #1 & #2 \\ #4 & #5 \end{smallmatrix} \big|
\begin{smallmatrix} #3 \\ #6 \end{smallmatrix}
\Big)\,}}

\newcommand{\gO}{\ensuremath{\op{g}}}

\newcommand{\qO}{\ensuremath{\op{q}}}
\newcommand{\rO}{\ensuremath{\op{r}}}

\newcommand{\CO}{\ensuremath{\op{C}}}

\newcommand{\HO}{\ensuremath{\op{H}}}

\newcommand{\TO}{\ensuremath{\op{T}}}

\newcommand{\VO}{\ensuremath{\op{V}}}

\newcommand{\Vnnn}{\ensuremath{\op{V}_\text{3N}}}

\newcommand{\pOV}{\ensuremath{\vec{\op{p}}}}
\newcommand{\qOV}{\ensuremath{\vec{\op{q}}}}
\newcommand{\rOV}{\ensuremath{\vec{\op{r}}}}

\newcommand{\sigmaOV}{\ensuremath{\vec{\op{\sigma}}}}

\newcommand{\cm}{\ensuremath{\textrm{cm}}}
\newcommand{\nuc}[2]{\ensuremath{{}^{#2}\text{#1}}}

\newcommand{\fm}{\ensuremath{\,\text{fm}}}

\newcommand{\MeV}{\ensuremath{\,\text{MeV}}}

\newcommand{\Cnnn}{\ensuremath{C_\text{3N}}}
\newcommand{\ennn}{\ensuremath{e_\text{3N}}}

\newcommand{\symboldiamond}[1][black]{{\color{#1}\scriptsize\begin{turn}{45}$\blacksquare$\end{turn}}}
\newcommand{\symboltriangle}[1][black]{{\color{#1}$\blacktriangle$}}
\newcommand{\symbolbox}[1][black]{{\color{#1}\small$\blacksquare$}}
\newcommand{\symbolcircle}[1][black]{{\color{#1}$\bullet$}}

\definecolor{FGViolet}{rgb}{0.61,0.32,0.61}
\definecolor{FGDarkBlue}{rgb}{0,0,0.6}
\definecolor{FGBlue}{rgb}{0,0,0.8}
\definecolor{FGLightBlue}{rgb}{0.2, 0.6, 0.8}
\definecolor{FGGreen}{rgb}{0.2,0.7,0.2}
\definecolor{FGLightGreen}{rgb}{0.4,1,0.4}
\definecolor{FGYellow}{rgb}{1,0.95,0}
\definecolor{FGDarkYellow}{rgb}{1,0.8,0.3}
\definecolor{FGOrange}{rgb}{0.95,0.5,0.1}
\definecolor{FGRed}{rgb}{0.8,0,0}
\definecolor{FGWhite}{rgb}{1,1,1}
\definecolor{FGLightGray}{rgb}{0.8,0.8,0.8}
\definecolor{FGGray}{rgb}{0.5,0.5,0.5}
\definecolor{FGDarkGray}{rgb}{0.3,0.3,0.3}
\definecolor{FGBlack}{rgb}{0,0,0}

\newcommand{\linemediumsolid}[1][black]{\unitlength 0.7ex
  {\color{#1}
  \begin{picture}(6,1)
  \linethickness{0.4mm}
  \put(0,0.5){\line(1,0){6.0}}
  \end{picture}}\nolinebreak
}

\newcommand{\linemediumdashed}[1][black]{\unitlength 0.7ex
  {\color{#1}
  \begin{picture}(6,1)
  \linethickness{0.4mm}
  \put(0,0.5){\line(1,0){1.5}}
  \put(2.2,0.5){\line(1,0){1.5}}
  \put(4.4,0.5){\line(1,0){1.5}}
  \end{picture}}\nolinebreak
}

\begin{document}

\title{Systematics of binding energies and radii based on realistic two-nucleon \\plus phenomenological three-nucleon interactions}

\author{A. G\"unther}
\email{anneke.guenther@physik.tu-darmstadt.de}
\affiliation{Institut f\"ur Kernphysik, Technische Universit\"at Darmstadt,
64289 Darmstadt, Germany}

\author{R. Roth}
\email{robert.roth@physik.tu-darmstadt.de}
\affiliation{Institut f\"ur Kernphysik, Technische Universit\"at Darmstadt,
64289 Darmstadt, Germany}

\author{H. Hergert}
\affiliation{National Superconducting Cyclotron Laboratory, Michigan State University, East Lansing, MI 48824, USA}

\author{S. Reinhardt}
\affiliation{Institut f\"ur Kernphysik, Technische Universit\"at Darmstadt,
64289 Darmstadt, Germany}

\date{\today}

\begin{abstract}
We investigate the influence of phenomenological three-nucleon interactions on the systematics of ground-state energies and charge radii throughout the whole nuclear mass range from \nuc{He}{4} to \nuc{Pb}{208}. The three-nucleon interactions supplement unitarily transformed two-body interactions constructed within the Unitary Correlation Operator Method or the Similarity Renormalization Group approach. To be able to address heavy nuclei as well, we treat the many-body problem in Hartree-Fock plus many-body perturbation theory, which is sufficient to assess the systematics of energies and radii, and limit ourselves to regularized three-body contact interactions. We show that even with such a simplistic three-nucleon interaction a simultaneous reproduction of the experimental ground-state energies and charge radii can be achieved, which is not possible with unitarily transformed two-body interactions alone.  
\end{abstract}


\maketitle

\clearpage

\section{Introduction}

Nuclear structure theory is approaching an era of systematic many-body calculations using nuclear Hamiltonians based on Quantum Chromodynamics (QCD). An important step along these lines is the formulation of nuclear interactions within chiral effective field theory \cite{Entem:2003ft,Epelbaum:2002vt,Epelbaum:2005pn}, leading to a consistent hierarchy of two-, three- and many-nucleon interactions starting from the relevant degrees of freedom and symmetries for the low-energy nuclear structure regime. The use of these two-, three- and many-nucleon interactions in nuclear structure calculations is a formidable task. 

In addition to few-body calculations the most promising nuclear structure calculations using the chiral two- plus three-nucleon interaction consistently have been performed in the no-core shell model (NCSM)  for mid p-shell nuclei \cite{Navratil:2007xxx}. An immense numerical effort is needed to compute and manage the three-body matrix elements in these calculations, which limits the range of applicability of these calculations at present. Recently, the use of consistent two- plus three-nucleon interactions resulting from a Similarity Renormalization Group evolution of the chiral two- plus three-nucleon interaction was demonstrated also in the context of the NCSM \cite{Jurgenson:2009qs}. This approach, a unitary transformation of the chiral Hamiltonian aiming at a pre-diagonalization that improves the convergence properties of NCSM substantially, holds great potential also for the use in other many-body  schemes and will play a significant role in the future. However, the computational effort for including those two- plus three-nucleon interactions into  many-body calculations, be it exact or approximate, is still the limiting factor for many applications.

In this paper we follow a more pragmatic route to explore the impact of three-body forces in connection with unitarily transformed two-nucleon interactions. We start from the Argonne V18 high-precision two-nucleon potential \cite{Wiringa:1994wb}, which is still widely used although it does not have the same systematic link to QCD like the chiral effective field theory interactions and is considered phenomenological in this respect. We then use the Similarity Renormalization Group \cite{Bogner:2006pc,Hergert:2007wp,Roth:2008km,Roth:2009ppnp,Bogner:2010ppnp} as well as the Unitary Correlation Operator Method \cite{Feldmeier:1997zh,Neff:2002nu,Roth:2004ua,Roth:2009ppnp} to construct a transformed two-nucleon interaction, which has a much better convergence behavior and allow us to use simplified many-body schemes. At this level neither genuine nor induced three-nucleon interactions are included. From various applications of these unitarily transformed two-nucleon interactions we know that there are characteristic deviations of basic nuclear observables from the experimental systematics that might be connected to three-body interactions. For example, unitarily transformed two-body interactions which yield a realistic systematics for binding energies tend to underestimate the charge radii \cite{Roth:2005ah,Roth:2009ppnp}. Here we study to what extend these systematic deviations can be cured by including a three-body interaction. Note that we are not aiming at a precision description of individual nuclei but rather the complete systematics from light nuclei, \nuc{He}{4}, to heavy nuclei, \nuc{Pb}{208}.

To facilitate calculations for the full mass range from \nuc{He}{4} to \nuc{Pb}{208} we have to simplify the approach compared to the consistent. The first simplification consists in the use of a phenomenological three-body interaction, which allows for an efficient computation of matrix elements but violates the consistency discussed above. The second simplification consists in the use of Hartree-Fock plus many-body perturbation theory for the approximate solution of the many-body problem. Despite of these simplifications, we will obtain valuable information on the interplay between realistic two-body and phenomenological three-body interactions and on how well the experimental systematics of ground-state energies and charge radii can be reproduced. Furthermore, these studies prepare the ground for calculations with consistently transformed two- plus three-nucleon interactions. 

After a brief reminder of the basic concepts of the Unitary Correlation Operator Method and the Similarity Renormalization Group we introduce the phenomenological three-body interaction and calculate the matrix elements in the harmonic oscillator basis in the second section. In the third section, we discuss the inclusion of the three-body interaction in Hartree-Fock and many-body perturbation theory and discuss the systematics of ground-state energies and charge rms-radii across the whole nuclear mass range and its dependence on the two- and three-nucleon interaction.

\section{formalism}\label{sec:formal}

\subsection{Unitary Correlation Operator Method and Similarity Renormalization Group}\label{ssec:ucom+srg}

The Unitary Correlation Operator Method and the Similarity Renormalization Group provide two conceptually different but physically related approaches for the construction of soft phase-shift equivalent interactions. 

The Similarity Renormalization Group (SRG) \cite{Bogner:2006pc,Hergert:2007wp,Roth:2008km,Roth:2009ppnp,Bogner:2010ppnp} aims at the pre-diagonalization of the Hamiltonian for a given basis by means of a unitary transformation implemented through the renormalization-group flow equation:
\begin{equation}\label{eq:srgevolution}
 \frac{d\HO_\alpha}{d\alpha}=[\eta_\alpha,\HO_\alpha] \;,
\end{equation}
where $\alpha$ is the flow parameter and $\HO_\alpha$ the evolved Hamiltonian, with $\HO_0=\HO$ being the initial or `bare' Hamiltonian. The anti-hermitian generator $\eta_{\alpha}$ defines the specifics of the flow evolution, e.g. the representation with respect to which the Hamiltonian should become diagonal or block-diagonal. Various choices for this generator have been investigated \cite{Bogner:2010ppnp}, we restrict ourselves to the simple generator \cite{Roth:2008km, Bogner:2006pc}
\begin{equation}\label{eq:srggenerator}
 \eta_\alpha=[\TO_{\text{int}},\HO_\alpha]
\end{equation}
with $\TO_{\text{int}} = \TO-\TO_{\cm}$ being the intrinsic kinetic energy, which leads to a pre-diagonalization of the Hamiltonian with respect to the eigenbasis of the kinetic energy or momentum operator. Once the generator is fixed, the Hamiltonian and all operators of interest can be evolved easily using a matrix representation of the flow equation \eqref{eq:srgevolution}. 

In $A$-body space the evolution generates up to $A$-body operators even if the initial Hamiltonian contains only up to two- or three-body operators. For reasons of practicability one has to truncate the evolution at some low particle number---typically this is done by solving the evolution equation in a matrix representation in two- or three-body space. For the moment we restrict ourselves to transformations in two-body space, i.e., we will discard any induced three-body interactions.

The aim of the Unitary Correlation Operator Method (UCOM) \cite{Feldmeier:1997zh,Neff:2002nu,Roth:2004ua,Roth:2005pd,Roth:2009ppnp} is to explicitly treat short-range correlations induced by the nuclear interaction via a static unitary transformation. This transformation can either be used to correlate the many-body states or to similarity transform operators of interest, e.g. the Hamiltonian
\begin{equation}
  \tilde{\HO}=\CO^{\dagger}\HO\CO \;,
\end{equation}
using the correlation operator $\CO$. The dominant short-range correlations are induced by the strong short-range repulsion and the tensor part of the nuclear interaction. Therefore the correlation operator is written as a product of two unitary operators, $\CO_r$ for the central correlations and $\CO_{\Omega}$ for the tensor correlations. We choose an explicit form of the correlation operators:
\begin{equation}
 \CO=\CO_{\Omega}\CO_r=\text{exp}\Big{(}-i\sum_{j<k}\gO_{\Omega,jk}\Big{)}\ 
 \text{exp}\Big{(}-i\sum_{j<k}\gO_{r,jk}\Big{)}
\end{equation}
with the following ansatz for hermitian generators $\gO_r$ and $\gO_{\Omega}$:
\begin{equation}
\label{eq:ucomgenerators}
\begin{split}
\gO_r &= \tfrac{1}{2} [\qO_r s(\rO) + s(\rO) \qO_r] \;, \\
\gO_{\Omega} &= \tfrac{3}{2} [(\sigmaOV_1\cdot\rOV)(\sigmaOV_2\cdot\qOV_{\Omega}) 
                          + (\sigmaOV_1\cdot\qOV_{\Omega})(\sigmaOV_2\cdot\rOV)] \;,
\end{split}
\end{equation}
where $\qO_r = \frac{1}{2} (\frac{\rOV}{\rO}\cdot\qOV + \qOV\cdot\frac{\rOV}{\rO})$, $\qOV_{\Omega} = \qOV-\frac{\rOV}{\rO}\cdot\qO_r$, and $\qOV = \frac{1}{2}[\pOV_1 - \pOV_2]$. The strengths and radial dependencies of the two transformations are governed by the correlation functions $s(r)$ and $\vartheta(r)$ for the central and tensor correlations, respectively. One can obtain these functions via an energy minimization in the two-body system \cite{Roth:2005pd}. Recently, we have also employed the SRG as tool for the determination of the UCOM correlation functions $s(r)$ and $\vartheta(r)$ as discussed in Refs.~\cite{Roth:2008km, Roth:2009ppnp}. Here, we will use these SRG-optimized UCOM correlation functions only.

Though the SRG- and UCOM-transformations have a different formal background, they address the same physics of short-range correlations. A first connection becomes clear at the level of the generators \cite{Hergert:2007wp}---the SRG generator \eqref{eq:srggenerator} in two-body space at $\alpha=0$ reveals the same operator structures that appear in the UCOM-generators \eqref{eq:ucomgenerators}. At the level of matrix elements, both the SRG- and UCOM-transformations lead to a suppression of the off-diagonal momentum-space matrix elements and an enhancement of the low-momentum matrix elements as discussed in detail in Ref.~\cite{Roth:2009ppnp}. 

In the following, we employ both transformations to generate one-parameter families of phase-shift equivalent two-body interactions starting from a specific initial NN-interaction, the Argonne V18 (AV18) in our case. For the SRG-transformation the flow parameter $\alpha$ directly spans this family of two-body interactions. We will study two versions of the SRG-transformation, one where the flow equations are solved for all partial waves and one where only the partial-waves containing relative $S$-waves, i.e. the ${}^1S_0$ and the coupled ${}^3S_1-{}^3D_1$ partial waves, are transformed. The latter is motivated by the fact that short-range correlations affect the $S$-wave channels most, because for all higher orbital angular momenta the relative wave functions are suppressed by the centrifugal barrier at short distances. We use the label `SRG' for the fully transformed interactions and `S-SRG' for the S-wave-only transformations. For the UCOM-transformation we use correlation functions determined from SRG-evolved two-body wave functions as discussed in Refs. \cite{Roth:2008km, Roth:2009ppnp}, thus the flow parameter $\alpha$ also spans a family of different UCOM-transformed interactions. Note that the standard formulation of UCOM only uses different transformations for the different $(S,T)$-channels. We thus use the SRG-evolved wave functions for the lowest partial waves for each $(S,T)$-channel to define the correlation functions, leading to a transformed interaction labelled `UCOM(SRG)'. Analogously to the S-SRG transformation, we can also use an S-wave-only UCOM transformation, denoted `S-UCOM(SRG)', which acts only in the ${}^1S_0$ and the coupled ${}^3S_1-{}^3D_1$ partial waves. 

So far, we have assumed that both transformations are evaluated in two-body space, leading to a transformed interaction containing two-body terms only. A consistent first-principles treatment requires the transformation to be performed in $A$-body space, leading to a hierarchy of induced interactions up to the $A$-body level, as mentioned earlier. The most advanced attempts along these lines use the full SRG-evolution at the three-body level to construct a consistently transformed two- plus three-nucleon interaction \cite{Jurgenson:2009qs}. The use of those two- plus three-body interactions in many-body calculations is very demanding and presently limited to rather small model spaces. 

Therefore, we follow a more pragmatic path in this work. We evaluate the unitary transformations at the two-body level and mimic the three-body contributions (genuine plus induced) through a simple phenomenological three-body interaction. By using a simplified three-nucleon (3N) interaction, e.g., a regularized contact or a Gaussian interaction, the calculation of the three-body matrix elements becomes formally and computationally much less demanding. This allows us to study the impact of 3N interactions on various nuclear structure observables for nuclei and model spaces beyond the domain accessible with realistic 3N interactions. Furthermore, we can develop and benchmark approximate treatments of the three-body contributions and establish the technical framework to include 3N interaction into different many-body methods. 

\begin{figure}
\centering
\includegraphics*[width=0.7\columnwidth]{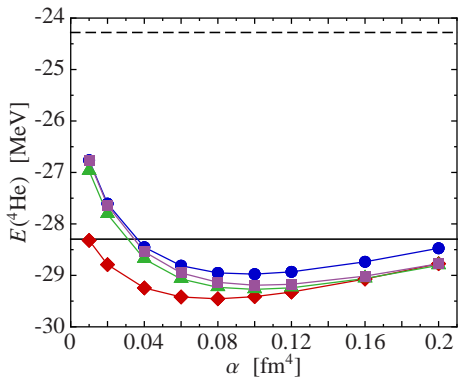}
\vspace{-0.3cm}
\caption{(color online) Binding energy of $\nuc{He}{4}$ as function of the flow parameter $\alpha$ obtained from a converged no-core shell model calculation using the UCOM(SRG)-transformed (\symbolcircle[FGBlue]), the S-UCOM(SRG)-transformed (\symboldiamond[FGRed]), the SRG-transformed (\symboltriangle[FGGreen]), or the S-SRG-transformed (\symbolbox[FGViolet]) AV18 potential. The horizontal lines indicate the experimental binding energy (\linemediumsolid[black]) and the exact ground state energy for the bare AV18 two-body interaction (\linemediumdashed[black]) \cite{Nogga:2000xxx}. 
}
\label{fig:ncsm_He4}
\end{figure}

The parameters of the phenomenological 3N interactions will be adjusted depending on the flow parameter $\alpha$ of the transformed two-nucleon (NN) interaction. For a wide range of $\alpha$ parameters the transformed two-body interaction alone produces an overbinding compared to the experimental ground state energy. This is illustrated in Fig. \ref{fig:ncsm_He4} for the ground-state energy of \nuc{He}{4} as function of $\alpha$ obtained in converged no-core shell model calculations for the UCOM(SRG)-, the S-UCOM(SRG)-, the SRG-, and the S-SRG-transformed AV18 interaction. Thus the additional phenomenological interaction, which mimics the net effect of the genuine and the induced 3N interaction, has to be repulsive in order to lead to a \nuc{He}{4} binding energy consistent with experiment. Note that the phenomenological three-body forces that are used in connection with the bare AV18 interaction are generally attractive. Thus the induced 3N interaction resulting from the unitary transformation of the NN interaction alone has to be repulsive and sufficiently strong to create an over-all repulsive three-body contribution.

\subsection{Three-Body Contact Interaction}
\label{ssec:3b}

The simplest choice for a phenomenological 3N interaction is a spin-isopin-independent contact interaction
\begin{equation}\label{eq:3bint}
  \Vnnn
  = \Cnnn\ \delta^{(3)}(\op{\vec{x}}_1-\op{\vec{x}}_2)\ \delta^{(3)}(\op{\vec{x}}_1-\op{\vec{x}}_3)
\end{equation}
with variable strength $\Cnnn$. Despite its simplicity it allows us to study the impact of a 3N interaction on bulk observables like ground-state energies or charge radii. Obviously this simplistic choice offers substantial computational advantages. 

For evaluating the matrix elements of a realistic 3N interaction for the use in configuration-space Hartree-Fock or no-core shell model type calculations one typically adopts a two-step procedure: First the matrix elements are evaluated in a Jacobi-coordinate basis for the relative motion in the three-nucleon system. Then, through a sequence of Talmi-Moshinski transformations and angular momentum recouplings, the matrix elements are transformed into the m-scheme to perform the many-body calculation. Both steps are non-trivial and computationally demanding, thus limiting the model-space sizes for which those matrix elements can be handled. 

In contrast, the matrix elements of the contact interaction can be directly evaluated in the m-scheme in a straight-forward manner. We first consider the matrix elements of the 3N contact interaction with respect to the spatial part of three-particle product states in the harmonic oscillator basis
\eq{
  \bra{n_1 l_1 m_{l_1},n_2 l_2 m_{l_2},n_3 l_3 m_{l_3}} \Vnnn 
  \ket{n_4 l_4 m_{l_4},n_5 l_5 m_{l_5},n_6 l_6 m_{l_6}} \;.
}
The spin and isospin quantum numbers and the antisymmetrization will be included subsequently. We can insert a unit operator in position representation using cartesian coordinates and directly evaluate the Kronecker-deltas. This leaves us with a single integration over a single-particle coordinate, which can be rewritten in spherical coordinates. Introducing the position representation of the harmonic oscillator single-particle states, $\phi_{nlm_l}(\vec{x}) = R_{nl}(x) Y_{lm_l}(\Omega)$, with radial wave functions $R_{nl}(x)$ and spherical harmonics $Y_{lm_l}(\Omega)$, we obtain:
\eqmulti{ \label{eq:matrixelem3N}
  &\bra{n_1 l_1 m_{l_1},n_2 l_2 m_{l_2},n_3 l_3 m_{l_3}} \Vnnn 
  \ket{n_4 l_4 m_{l_4},n_5 l_5 m_{l_5},n_6 l_6 m_{l_6}} \\
  &\phantom{mm} = \Cnnn \int dx x^2 R_{n_1 l_1}(x) R_{n_2 l_2}(x) R_{n_3 l_3}(x) \\
  &\phantom{mmmmmmmmm} \times R_{n_4 l_4}(x) R_{n_5 l_5}(x) R_{n_6 l_6}(x) \\
  &\phantom{mmm} \times \int d\Omega \ Y_{l_1 m_{l_1}}^*(\Omega) Y_{l_2 m_{l_2}}^*(\Omega) Y_{l_3 m_{l_3}}^*(\Omega) \\
  &\phantom{mmmmmmm} \times Y_{l_4 m_{l_4}}(\Omega) Y_{l_5 m_{l_5}}(\Omega) Y_{l_6 m_{l_6}}(\Omega) \ .
}
The integral over the six radial wave functions $R_{n l}(x)$ has to be calculated numerically while the integral over the six spherical harmonics $Y_{l m_l}(\Omega)$ can be evaluated analytically. The product of three spherical harmonics can be reduced to one spherical harmonic and the integral over the remaining two spherical harmonics can be solved analytically, leading to
\eqmulti{
\int d\Omega& Y_{l_1 m_{l_1}}^*(\Omega) Y_{l_2 m_{l_2}}^*(\Omega) Y_{l_3 m_{l_3}}^*(\Omega) Y_{l_4 m_{l_4}}(\Omega) Y_{l_5 m_{l_5}}(\Omega) Y_{l_6 m_{l_6}}(\Omega) \\
=& \frac{1}{16 \pi^2} \hat{l_1}\, \hat{l_2}\, \hat{l_3}\, \hat{l_4}\, \hat{l_5}\, \hat{l_6}\, 
\sum_{\substack{L_1 L_2 L_3 \\ M_{L_1} M_{L_2} M_{L_3}}}
\frac{1}{2L_2+1} \\
& \times
\clebsch{l_1}{l_2}{L_1}{0}{0}{0}
\clebsch{L_1}{l_3}{L_2}{0}{0}{0}
\clebsch{l_4}{l_5}{L_3}{0}{0}{0}
\clebsch{L_3}{l_6}{L_2}{0}{0}{0} \\
& \times
\clebsch{l_1}{l_2}{L_1}{m_{l_1}}{m_{l_2}}{M_{L_1}}
\clebsch{L_1}{l_3}{L_2}{M_{L_1}}{m_{l_3}}{M_{L_2}} \\
& \times
\clebsch{l_4}{l_5}{L_3}{m_{l_4}}{m_{l_5}}{M_{L_3}}
\clebsch{L_3}{l_6}{L_2}{M_{L_3}}{m_{l_6}}{M_{L_2}}
}
with $\hat{l}=\sqrt{2l+1}$ and $\clebsch{l_1}{l_2}{L}{m_{l_1}}{m_{l_2}}{M_{L}}$ being Clebsch-Gordan coefficients.

We precompute and store those angular integrals as well as the radial integrals in \eqref{eq:matrixelem3N}. The inclusion of the spin and isospin quantum numbers, the coupling of the single-particle orbital angular momenta and the spins, and the antisymmetrization are then done on the flight during the many-body calculation. This makes calculations in large model spaces feasible. 

For applications beyond the mean-field level a regularization of the contact interaction is inevitable. However, the regularization should preserve the simplicity of the matrix-element calculation, which rules out momentum-space cutoffs and such. Hence, we introduce an energy cut-off parameter $\ennn$, which defines an upper bound for total oscillator energy of the three-particle state, $(2n_1+l_1)+(2n_2+l_2)+(2n_3+l_3)\leq\ennn$. The implementation of this cutoff is trivial and it preserves all computational advantages of the contact interaction.

\section{Many-Body Calculations}
\label{sec:results}

We adopt the 3N contact interaction together unitarily transformed NN interactions for the study of the systematics of nuclear ground-state energies and charge radii throughout the whole mass range from \nuc{He}{4} to \nuc{Pb}{208} using Hartree-Fock and many-body perturbation theory.

\subsection{Hartree-Fock Approximation}
\label{ssec:hf}

We have employed the Hartree-Fock (HF) approximation as a first indicator for the gross systematics of binding energies and charge radii obtained with unitarily transformed two-body interactions in Refs. \cite{Roth:2005ah,Roth:2009ppnp} already. In order to assess the impact of 3N contact interactions we extend our HF framework in a first step. 

All calculations are based on the translationally invariant Hamiltonian
\eq{\label{eq:Hint}
  \op{H}_{\text{int}}
  =\TO_{\text{int}} + \VO_{\text{NN}}+\Vnnn 
  =\HO_{\text{int}}^{(2)}+\Vnnn
}
with $\VO_{\text{NN}}$ being the UCOM- or SRG-transformed NN interaction and $\TO_{\text{int}}=\TO-\TO_{\text{cm}}$ the intrinsic kinetic energy. This Hamiltonian includes all charge dependent and electromagnetic terms of the transformed AV18 potential as well as the phenomenological three-body force. 

The HF equations are formulated in a harmonic oscillator basis representation, i.e., the single-particle states are expanded in the harmonic oscillator states:
\begin{equation} \label{eq:hfstate}
\ket{\nu ljmm_t}=\sum_n C^{(\nu ljm_t)}_n \ket{nljmm_t} \;,
\end{equation}
where $\ket{nljmm_t}$ denotes the harmonic oscillator eigenstates with radial quantum number $n$, orbital angular momentum $l$, total angular momentum $j$ with projection $m$, and isospin projection quantum number $m_t$. Since we only consider closed-shell nuclei in the following, the expansion coefficients are independent of $m$. The HF equations can now be written as
\begin{equation} \label{eq:hfequations}
\sum_{\bar{n}}h^{(ljm_t)}_{n\bar{n}}C^{(\nu ljm_t)}_{\bar{n}}=\varepsilon^{(\nu ljm_t)}C^{(\nu ljm_t)}_n
\end{equation}
with the single-particle energies $\varepsilon^{(\nu ljm_t)}$. The matrix elements of the single-particle HF Hamiltonian
\eqmulti{
h^{(ljm_t)}_{n\bar{n}}
&= \sum_{l'j'm'_t}\sum_{n'\bar{n}'}  
   \matrixe{nljm_t,n'l'j'm'_t}{\HO_{\text{int}}^{(2)}}{\bar{n}ljm_t, \bar{n}'l'j'm'_t}\; \varrho^{(l'j'm'_t)}_{\bar{n}'n'} \\
&+ \frac{1}{2}\sum_{\substack{l'j'm'_t\\ l''j''m''_t}}\sum_{\substack{n'n''\\ \bar{n}'\bar{n}''}}
     \bra{nljm_t,n'l'j'm'_t,n''l''j''m''_t} \times \\
&\times \VO_{\text{3N}} \ket{\bar{n}ljm_t, \bar{n}'l'j'm'_t,\bar{n}''l''j''m''_t}\; 
     \varrho^{(l'j'm'_t)}_{\bar{n}'n'}\varrho^{(l''j''m''_t)}_{\bar{n}''n''} 
}
are obtained by contractions of the antisymmetrized matrix elements of the two-body part of the Hamiltonian $\op{H}_{\text{int}}^{(2)}$ and the three-body interaction $\VO_{\text{3N}}$ with the one-body density matrix given by
\begin{equation}
\varrho^{(ljm_t)}_{\bar{n}n}=\sum_{\nu}O^{(\nu ljm_t)}C^{(\nu ljm_t)^*}_{\bar{n}} C^{(\nu ljm_t)}_n
\end{equation}
with $O^{(\nu ljm_t)}$ being the number of occupied magnetic sublevels which is $2j+1$ for closed-shell nuclei.

In the following the HF approach is applied to selected closed-shell nuclei from \nuc{He}{4} to \nuc{Pb}{208}. The HF equations are solved iteratively until full self-consistency is reached. The model space is truncated at a given major oscillator quantum number $e=2n+l\leq e_{\max}$, where $e_{\max}=10$ is sufficient to obtain converged ground-state energies and radii at the HF level. The oscillator parameter is chosen for each nucleus separately such that the experimental charge radius is reproduced by a shell-model Slater determinant built from harmonic oscillator single-particle states.

\begin{figure}
\centering
\includegraphics*[width=\columnwidth]{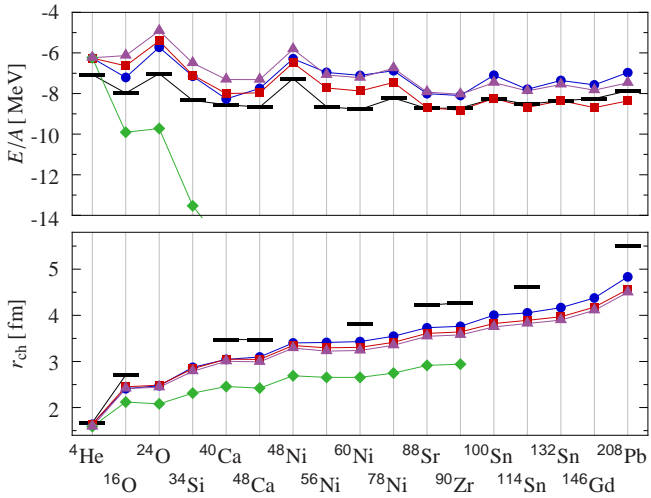}
\vspace{-0.3cm}
\caption{(color online) Ground-state energies per nucleon and charge radii of selected closed-shell nuclei resulting from HF calculations based on pure two-body interactions for $e_\text{max}=10$:
UCOM(SRG) with $\alpha=0.16\fm^4$ (\symbolcircle[FGBlue]),
S-UCOM(SRG) with $\alpha=0.16\fm^4$ (\symbolbox[FGRed]),
SRG with $\alpha=0.10\fm^4$ (\symboldiamond[FGGreen]),
S-SRG with $\alpha=0.10\fm^4$ (\symboltriangle[FGViolet]).
The bars indicate the experimental values \cite{Audi:1995dz,Vries:1987}.}
\label{fig:HF2bsrgD-srg}
\end{figure}

As a first illustration of the behavior of unitarily transformed two-body interactions Fig. \ref{fig:HF2bsrgD-srg} summarizes the ground-state energies per nucleon and the charge radii obtained at the HF level for nuclei up to \nuc{Pb}{208}. We adopt four different two-body interactions---UCOM(SRG), S-UCOM(SRG), SRG, and S-SRG---with flow parameters relevant for the later calculations including the 3N contact interaction. We observe that the general trend of the binding energies and charge radii is similar for the UCOM(SRG), the S-UCOM(SRG), and the S-SRG interactions. All three interactions produce binding energies that are within 2 MeV per nucleon of the experimental values for the whole mass range. By including correlations beyond HF, e.g., through many-body perturbation theory, all interactions would lead to an overbinding compared to experiment. At the same time the charge radii are underestimated for all but the lightest isotopes. Those systematic deviations can be remedied by a repulsive 3N interaction, as it will be included in the next step. 

The SRG-transformed interaction exhibits a vastly different behavior. The binding energies per nucleon increase rapidly with mass number, leading to an completely unphysical overbinding already at the HF level for intermediate and heavy nuclei. At the same time the charge radii are even smaller than the ones obtained with the other transformed interactions. Those strong systematic deviations have to be compensated by the 3N interaction that is generated from the initial NN potential in the course of the SRG-evolution. Because of the mere size of the three-body corrections needed one cannot expect a simple phenomenological interaction to be adequate to capture the main physics contained in the three-body contributions. Therefore, we will not consider the fully SRG-transformed interactions in the following.

\begin{figure}
\centering
\includegraphics*[width=\columnwidth]{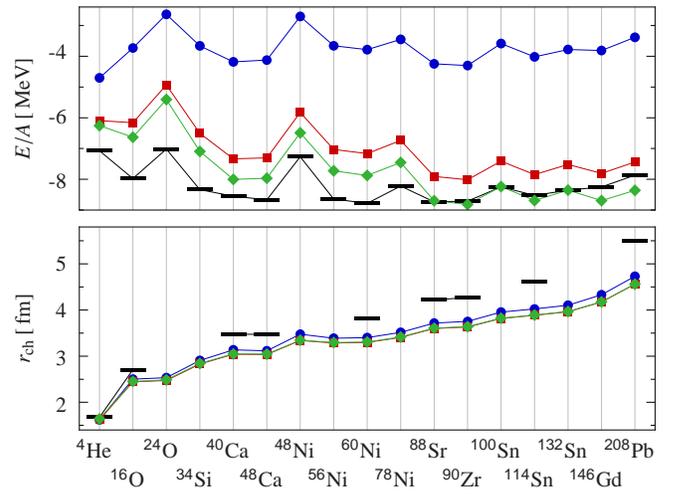}
\vspace{-0.3cm}
\caption{(color online) Ground-state energies per nucleon and charge radii of selected closed-shell nuclei resulting from HF calculations for the pure two-body interaction S-UCOM(SRG) for $e_\text{max}=10$ and different flow parameters:
$\alpha=0.04\fm^4$ (\symbolcircle[FGBlue]),
$\alpha=0.12\fm^4$ (\symbolbox[FGRed]),
$\alpha=0.16\fm^4$ (\symboldiamond[FGGreen]).
The bars indicate the experimental values \cite{Audi:1995dz,Vries:1987}.}
\label{fig:HF2b_alpha}
\end{figure}

Before including the 3N contact interaction explicitly, we analyze the dependence of the HF results obtained with the transformed two-body interactions on the flow parameter $\alpha$. In Fig.~\ref{fig:HF2b_alpha} the binding energies and charge radii for the S-UCOM(SRG) interactions with $\alpha=0.04\,\text{fm}^4$, $0.12\,\text{fm}^4$, and $0.16\,\text{fm}^4$ are shown. For the smallest flow parameter $\alpha=0.04\fm^4$ the ground-state energies reproduce the systematics of the experimental values up to a constant shift. The missing binding energy can be explained by beyond-HF correlations that can be recovered, e.g., by perturbation theory. This flow parameter would be used for calculations based on the pure NN interaction, as they are discussed in detail in Refs.~\cite{Roth:2009ppnp,Roth:2008km,Roth:2005ah}. 

When increasing the flow parameter entering into the construction of the S-UCOM(SRG) interaction to $\alpha=0.12\,\text{fm}^4$ or $0.16\,\text{fm}^4$ the ground-state energy at the HF level decreases substantially. For most nuclei the binding energy per nucleon more than doubles when going from $\alpha=0.04\,\text{fm}^4$ to $0.16\,\text{fm}^4$. For heavier nuclei the increase is larger, thus leading to a tilt of the ground-state energy systematics with respect to the experimental behavior. Unlike the energies, the charge radii exhibit a very weak $\alpha$-dependence as shown in the lower panel of Fig.~\ref{fig:HF2b_alpha}. For all flow parameters considered here, the radii are somewhat underestimated. The situation is very similar for the UCOM(SRG) and the S-SRG interactions.

This general phenomenology of ground-state energies and charge radii obtained from unitarily transformed interactions at larger flow parameters illustrates that the purely repulsive phenomenological 3N interaction can be used to improve energies and radii. Since the radii are insensitive to the flow parameter in a certain regime, we can fix the strength of the 3N interaction such that the systematics of the charge radii is in good agreement with experiment. The flow parameter can then be chosen to provide an optimal description of the ground-state energies in a beyond-HF calculation. 

\begin{figure}
\centering
\includegraphics*[width=\columnwidth]{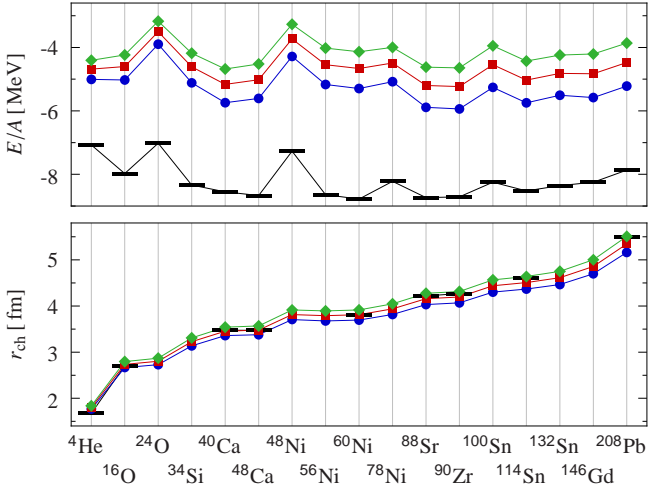}
\vspace{-0.3cm}
\caption{(color online) Binding energies per nucleon and charge radii of selected closed-shell nuclei resulting from HF calculations based on the S-UCOM(SRG) interaction for $\alpha=0.16\fm^4$, $e_\text{max}=10$, $\ennn=20$ and different strengths of the three-body interaction:
$\Cnnn=1600\MeV\fm^6$ (\symbolcircle[FGBlue]),
$\Cnnn=2200\MeV\fm^6$ (\symbolbox[FGRed]),
$\Cnnn=2800\MeV\fm^6$ (\symboldiamond[FGGreen]).
The bars indicate the experimental values \cite{Audi:1995dz,Vries:1987}.}
\label{fig:HFsrgD_C3N}
\end{figure}

The impact of 3N contact interactions with different strength parameters $C_{\text{3N}}$ is illustrated in Fig.~\ref{fig:HFsrgD_C3N} using the S-UCOM(SRG) interaction for $\alpha=0.16\,\text{fm}^4$. As compared to the HF calculation with the pure two-body interaction, the binding energies are reduced significantly and the charge radii are increased as a result of the purely repulsive 3N interaction. It is remarkable, that the charge radii are in excellent agreement with experiment for the whole mass range from \nuc{He}{4} to \nuc{Pb}{208} when using a 3N interaction with strength parameters in the range $C_{\text{3N}}=2200$ to $2800\,\text{MeV\,fm}^6$. For the same values $C_{\text{3N}}$ the ground-state energy systematics at the HF does again resemble the experimental systematics up to a constant shift, i.e. the tilt of the energy curve towards an overbinding for heavier nuclei is cured as well. The missing binding energy of $3$ to $4$ MeV per nucleon at the HF level can be recovered by including correlations beyond HF, as will be discussed in the next section.

\begin{figure}
\centering
\includegraphics*[width=0.85\columnwidth]{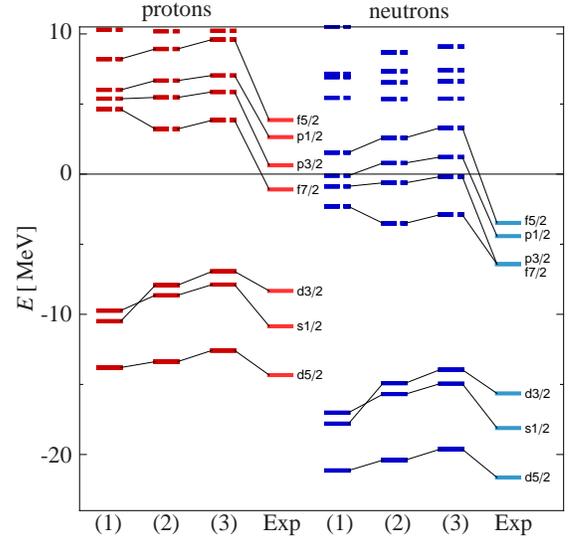}
\vspace{-0.3cm}
\caption{(color online) Single-particle spectra of $\nuc{Ca}{40}$ for different interactions: (1) UCOM(SRG) with $\alpha=0.16\fm^4$, $\Cnnn=1600\MeV\fm^6$, (2) S-UCOM(SRG) with $\alpha=0.16\fm^4$, $\Cnnn=2200\MeV\fm^6$, (3) S-SRG with $\alpha=0.10\fm^4$, $\Cnnn=2000\MeV\fm^6$. Three-body cut-off set to $\ennn=20$ for all calculations. Occupied states are indicated by solid lines, unoccupied states by dashed lines. Experimental data taken from Ref. \cite{Isakov:2002jv}.}
\label{fig:spsca}
\end{figure}

\begin{figure}
\centering
\includegraphics*[width=0.85\columnwidth]{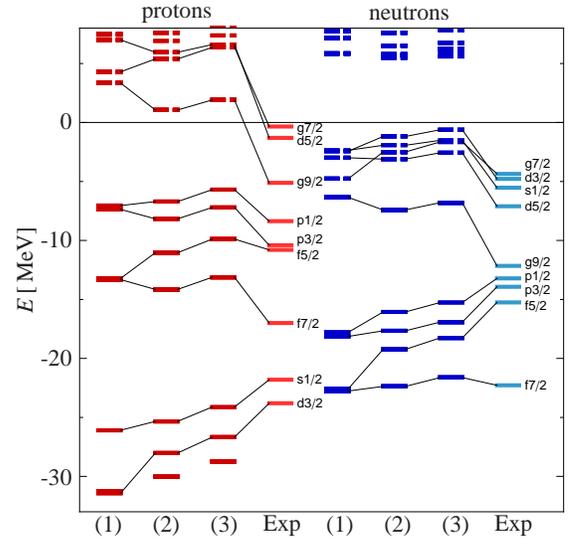}
\vspace{-0.3cm}
\caption{(color online) Single-particle spectra of $\nuc{Zr}{90}$ for the same interactions used in Fig. \ref{fig:spsca}. Experimental data taken from Refs. \cite{Wang:1993zzb,Delaroche:1989zz}.}
\label{fig:spszr}
\end{figure}

In addition to ground-state energies and radii the HF approximation provides us with an estimate for the single-particle energies that can be used to diagnose the various NN+3N interactions. Examples for the single particle spectra obtained with the various unitarily transformed interactions for \nuc{Ca}{40} and \nuc{Zr}{90} are shown in Figs. \ref{fig:spsca} and \ref{fig:spszr}, respectively. We use the UCOM(SRG) and the S-UCOM(SRG) interactions for $\alpha=0.16\,\fm^4$ and the S-SRG interaction for $\alpha=0.10\,\fm^4$ each supplemented with a 3N contact interaction with strength parameter adjusted to provide a good over-all description of the charge radii at the HF level, i.e., $\Cnnn=1600\MeV\fm^6$ for UCOM(SRG), $\Cnnn=2200\MeV\fm^6$ for S-UCOM(SRG), and $\Cnnn=2000\MeV\fm^6$ for S-SRG. 

The gross structure of the single-particle spectra obtained with the S-UCOM(SRG)+3N and the S-SRG+3N interactions agrees rather well with the single-particle energies extracted from experiment. The quality of the agreement is comparable with other mean-field type calculations and some of the of the characteristic deviations, e.g. the overestimation of the gaps at the Fermi energy, are expected to be remedied by the inclusion of beyond HF corrections. Other important quantities, e.g., the splittings between spin-orbit partner states, are reproduced rather well at the HF level already. 

The picture is different for the UCOM(SRG)+3N interaction. In particular for the single-particle spectrum of \nuc{Zr}{90} shown in Fig. \ref{fig:spszr} and for all heavier nuclei we observe a collapse of the spin-orbit splittings. Since this problem does not appear in the corresponding S-UCOM(SRG) calculation, it has to be caused by the UCOM-transformation of the higher partial waves. The problem is also absent in UCOM(SRG) interaction for smaller flow parameters $\alpha$, e.g. for the UCOM(SRG) interaction at $\alpha=0.04\fm^4$ that was used in Fig.~\ref{fig:HF2b_alpha}. Thus the long-range character of the tensor correlation functions as they appear for larger $\alpha$ (cf. Refs.~\cite{Roth:2008km,Roth:2009ppnp}) acting on the higher partial wave leads to this unphysical behavior. We will, therefore, restrict ourselves for the following discussion to the S-UCOM(SRG) and S-SRG interactions.

\subsection{Many-Body Perturbation Theory}
\label{ssec:mbpt}

A simple means to estimate the impact of correlations beyond the HF approximation is many-body perturbation theory (MBPT). In particular low-order MBPT corrections to the energy \cite{Goldstone:1957,Feldmeier:1977,Stevenson:2000cy,Coraggio:2003bd} can be computed quite efficiently for the whole mass range up to \nuc{Pb}{208}. We have used second- and third-order MBPT to investigate various two-body Hamiltonians and the importance and systematics of correlations beyond HF in Refs. \cite{Roth:2005ah,Roth:2009ppnp}. One should be aware, however, that low-order MBPT can only provide an estimate for the exact ground-state energies and that the order-by-order convergence is not guaranteed, as we have shown in Ref. \cite{Roth:2009up} using an harmonic oscillator single-particle basis. 

Because of its computational simplicity we adopt second-order MBPT as a guideline for the effect of beyond-HF correlations on the energy in the presence of a phenomenological 3N interaction. The second-order energy correction to the HF ground state energy for the intrinsic Hamiltonian \eqref{eq:Hint} including the 3N interaction reads
\eqmulti{ \label{eq:mbptenergy}
E^{(2)} 
&= \frac{1}{4}\sum^{<\varepsilon_F}_{hh'}\sum^{>\varepsilon_F}_{pp'}
\frac{\Big|\ \bra{hh'}\op{H}^{(2)}_{\text{int}}\ket{pp'}\ +
\sum\limits^{<\varepsilon_F}_{\bar{h}}\ \bra{hh'\bar{h}}\Vnnn\ket{pp'\bar{h}}\ \Big|^2 }{\varepsilon_{h}+\varepsilon_{h'}-\varepsilon_{p}-\varepsilon_{p'}} \\
&+ \frac{1}{36}\sum^{<\varepsilon_F}_{hh'h''}\sum^{>\varepsilon_F}_{pp'p''}
\frac{|\ \bra{hh'h''}\Vnnn\ket{pp'p''}\ |^2}
{\varepsilon_{h}+\varepsilon_{h'}+\varepsilon_{h''}
-\varepsilon_{p}-\varepsilon_{p'}-\varepsilon_{p''}} \;, \\
}
where $h,h',...$ denote the HF single-particle states \eqref{eq:hfstate} below the Fermi energy $\epsilon_F$ (hole states) and $p,p',...$ the corresponding HF single-particle states above the Fermi energy (particle states). All two- and three-body matrix elements appearing here are understood to be antisymmetrized matrix elements. 
 
Already the structure of the second-order energy correction \eqref{eq:mbptenergy} is interesting. Obviously, if we set all matrix elements of the 3N interaction to zero we recover the well known form of the second-order correction for a pure two-body Hamiltonian. The inclusion of the 3N interaction affects this expression in two ways: (\emph{i}) The matrix elements of the two-body Hamiltonian are modified by an effective or in-medium two-body term that results from the three-body matrix elements by a contraction of the third single-particle index. (\emph{ii}) An additional pure three-body term involving three particle and three hole indices appears. 

To separate the effect of these two contributions we study three variants of the second-order energy correction: MBPT(2B) includes only the contribution of the two-body Hamiltonian, i.e., the first matrix element in Eq.~\eqref{eq:mbptenergy}. MBPT(2B+3Bpphh) includes the in-medium two-body contribution generated by the 3N interaction, i.e., the complete first term in Eq.~\eqref{eq:mbptenergy}. Finally, MBPT(2B+3B) includes all terms of Eq.~\eqref{eq:mbptenergy}.

\begin{figure}
\centering
\includegraphics*[width=\columnwidth]{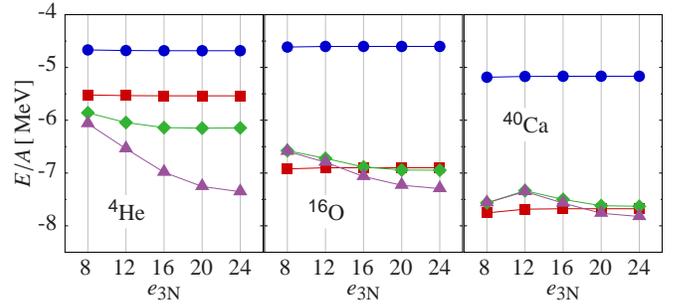}
\vspace{-0.3cm}
\caption{(color online) Contributions to the ground state energy resulting from MBPT based on the S-UCOM(SRG) interaction for $\alpha=0.16\fm^4$, $e_{\text{max}}=10$, $\Cnnn=2200\MeV\fm^6$ as function of the cut-off parameter $\ennn$ for 
HF (\symbolcircle[FGBlue]),
HF+MBPT(2B) (\symbolbox[FGRed]) ,
HF+MBPT(2B+3Bpphh) (\symboldiamond[FGGreen]), and
HF+MBPT(2B+3B) (\symboltriangle[FGViolet]).}
\label{fig:PTEe3N}
\end{figure}

The ground-state energies of \nuc{He}{4}, \nuc{O}{16}, and \nuc{Ca}{40} obtained with MBPT(2B), MBPT(2B+3Bpphh), and MBPT(2B+3B) on top of the HF result for the S-UCOM(SRG) interaction are shown in Fig. \ref{fig:PTEe3N} as function of the cutoff parameter $\ennn$. For the HF calculations presented so far we had fixed this cutoff to $\ennn=20$, which was sufficiently large to guarantee that the HF energies were practically independent of this cutoff for all nuclei. As soon as we include the second-order perturbative correction we cannot expect the results to be independent of $\ennn$, because the sums over particle states above the Fermi energy directly probe high-lying matrix elements. Eventually we will have to fix $\ennn$ to a certain value as part of the definition of the phenomenological 3N interaction. 

For the study of the different contributions a cutoff variation nevertheless provides a useful diagnostic tool. As seen in Fig.~\ref{fig:PTEe3N}, the HF energies are practically independent of $\ennn$ as mentioned earlier. When including the second-order correction due to the two-body Hamiltonian, MBPT(2B), the ground-state energies are lowered by about 1 MeV per nucleon for \nuc{He}{4} and by about 2.5 MeV per nucleon for \nuc{O}{16} and \nuc{Ca}{40}. The MBPT(2B) energies are sensitive to $\ennn$ only indirectly via high-lying HF single-particle states, therefore, the dependence is marginal. For MBPT(2B+3Bpphh) and MBPT(2B+3B) the cutoff directly affects the perturbative correction via the three-body matrix elements and the $\ennn$-dependence becomes more pronounced. Generally, the step from MBPT(2B) to MBPT(2B+3Bpphh) can modify the ground-state energy in either direction, whereas the change from MBPT(2B+3Bpphh) to MBPT(2B+3B) always results in a lowering of the ground-state energy, as evident from Eq. \eqref{eq:mbptenergy}. 

For \nuc{He}{4} we obtain a significantly lower ground-state energy when fully including the three-body terms. For the heavier nuclei the ground-state energy is increased at small $\ennn$ and remains almost unchanged for larger $\ennn$. Generally, the change in the ground-state energy per nucleon when going from MBPT(2B) to MBPT(2B+3B) for fixed and sufficiently large $\ennn$ decreases with increasing particle number. Beyond \nuc{Ca}{40} the impact of the three-body terms to the second-order energy correction is smaller than other uncertainties of the calculation, e.g., the degree of convergence with respect to the model-space. Therefore, we will limit ourselves to the MBPT(2B) corrections and will continue using $\ennn=20$ in the following. One should keep in mind, however, that for nuclei below \nuc{Ca}{40} and in particular for \nuc{He}{4} the full second-order correction MBPT(2B+3B) leads to a lower ground-state energy than MBPT(2B) and thus to a much better agreement with experiment.

\begin{figure}
\centering
\includegraphics*[width=\columnwidth]{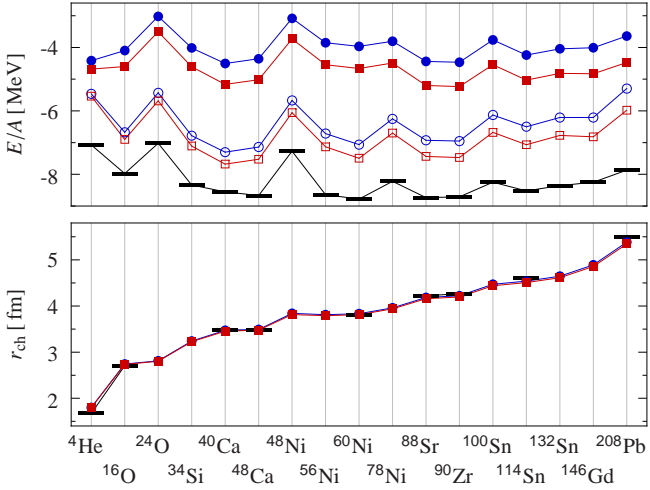}
\vspace{-0.3cm}
\caption{(color online) Binding energies per nucleon and charge radii of selected closed-shell nuclei resulting from HF calculations (filled symbols) and MBPT (open symbols) based on the S-UCOM(SRG) interaction for $e_\text{max}=10$, $\Cnnn=2200\MeV\fm^6$, $\ennn=20$ and different flow parameters:
$\alpha=0.12\fm^4$ (\symbolcircle[FGBlue]),
$\alpha=0.16\fm^4$ (\symbolbox[FGRed]).
The bars indicate the experimental values \cite{Audi:1995dz,Vries:1987}.}
\label{fig:HFsrgD}
\end{figure}

As discussed earlier, we can fix the strength of the three-body interaction based on the systematics of the charge radii and use the flow parameter entering into the two-body interaction to control the binding-energy systematics. In Fig. \ref{fig:HFsrgD} we illustrate the influence of $\alpha$ on the energies obtained in MBPT(2B) using the S-UCOM(SRG) interaction with $\alpha=0.12\fm^4$ and $0.16\fm^4$ and a 3N contact interaction with $\Cnnn=2200\MeV\fm^6$. Whereas the charge radii are practically identical for both values of $\alpha$, the HF and the MBPT(2B) ground state energies are systematically lower for the larger flow parameters. The difference is smaller for the MBPT(2B) energies than for the HF energies, as to be expected. The unitary transformation for larger $\alpha$ accounts for more of the correlations explicitly, thus the ground-state energy at the HF level is lower and the gain due to the inclusion of residual correlations through MBPT(2B) is smaller. Thus we can still use the $\alpha$-dependence to control the binding-energy systematics.

\begin{figure}
\centering
\includegraphics*[width=\columnwidth]{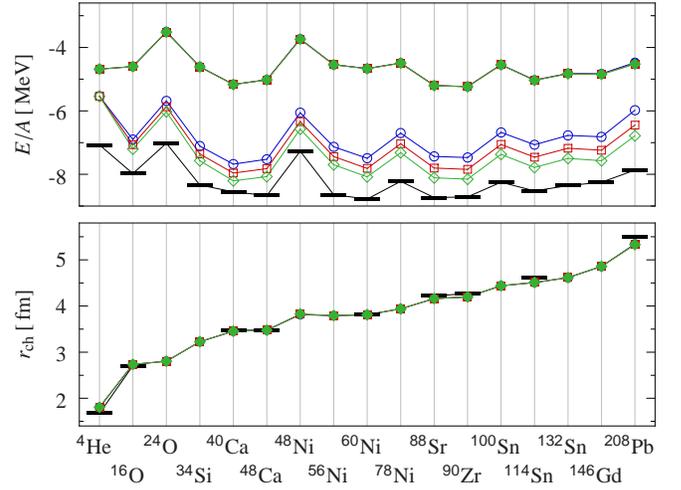}
\vspace{-0.3cm}
\caption{(color online) Binding energies per nucleon and charge radii of selected closed-shell nuclei resulting from HF calculations (filled symbols) and MBPT (open symbols) based on the S-UCOM(SRG) interaction for $\alpha=0.16\fm^4$, $\Cnnn=2200\MeV\fm^6$, $\ennn=20$ and different basis sizes:
$e_{\text{max}}=10$ (\symbolcircle[FGBlue]);
$e_{\text{max}}=12,\ l_{\text{max}}=10$ (\symbolbox[FGRed]);
$e_{\text{max}}=14,\ l_{\text{max}}=10$ (\symboltriangle[FGGreen]).
The bars indicate the experimental values \cite{Audi:1995dz,Vries:1987}.}
\label{fig:PTEsrgD}
\end{figure}

\begin{figure}
\centering
\includegraphics*[width=\columnwidth]{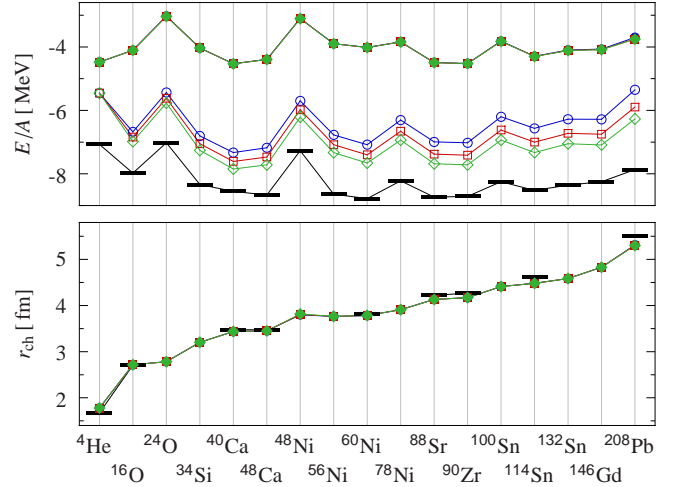}
\vspace{-0.3cm}
\caption{(color online) Binding energies per nucleon and charge radii of selected closed-shell nuclei resulting from HF calculations (filled symbols) and MBPT (open symbols) based on the S-SRG interaction for $\alpha=0.10\fm^4$, $\Cnnn=2000\MeV\fm^6$, $\ennn=20$ and different basis sizes:
$e_{\text{max}}=10$ (\symbolcircle[FGBlue]);
$e_{\text{max}}=12,\ l_{\text{max}}=10$ (\symbolbox[FGRed]);
$e_{\text{max}}=14,\ l_{\text{max}}=10$ (\symboltriangle[FGGreen]).
The bars indicate the experimental values \cite{Audi:1995dz,Vries:1987}.}
\label{fig:PTEsrg}
\end{figure}

So far we have used a model space with $e_{\max}=10$ for all calculations. This is absolutely sufficient to obtain converged HF results, but it is not enough to obtain converged HF+MBPT energies. The convergence behavior of the HF+MBPT(2B) energies is illustrated in Fig. \ref{fig:PTEsrgD} for the S-UCOM(SRG) interaction with $\alpha=0.16\fm^4$ and in Fig. \ref{fig:PTEsrg} for the S-SRG interaction with $\alpha=0.10\fm^4$ using $e_{\max}=10, 12,$ and $14$. The HF energies and charge radii are fully converged in all cases, but not the HF+MBPT(2B) energies. Although the largest model space includes 15 major oscillator shells, this is still not sufficient to obtain convergence of the MBPT(2B) contribution for heavier nuclei. The change of energy per nucleon between successive model-space sizes increases with increasing nucleon number.  

The slow convergence is partly due to the use of S-wave only UCOM and SRG transformations for the construction of the two-nucleon interactions. For all higher-partial waves the transformed interactions are thus identical to the initial interaction. Since those partial waves become increasingly important for heavier nuclei, the deterioration of the convergence is unsurprising. The use of a softer all-channel transformed interactions, e.g. the standard SRG-interaction discussed in Fig.~\ref{fig:HF2bsrgD-srg}, would help with the convergence. However, for those interactions a simple contact force is not sufficient to provide reasonable energy and radius systematics.   

Despite the non-optimal convergence, the results presented in Figs.~\ref{fig:PTEsrgD} and \ref{fig:PTEsrg} show that after inclusion of beyond-HF correlations also the systematics of the ground-state energy is reproduced quite well. Given the additional gain in binding energy that is expected until convergence, as estimated from an extrapolation $e_{\max}\to\infty$, for heavier nuclei and the additional binding resulting from the three-body contributions to the second-order correction for light isotopes the energies are in good systematic agreement with experiment. The charge radii follow the experimental results very closely already at the HF level. Perturbative corrections to the radii, as studied in Ref. \cite{Roth:2005pd}, are very small and will not affect the general agreement.

\section{Conclusions \& Outlook}

We have investigated the systematics of binding energies and charge radii for closed-shell nuclei from \nuc{He}{4} to \nuc{Pb}{208} starting from unitarily transformed realistic NN interactions supplemented by phenomenological 3N forces. We have shown that already a simplistic 3N contact interaction is sufficient to cure the systematic deviations from experiment that the UCOM- and SRG-transformed two-body interactions exhibit. By supplementing an S-UCOM or S-SRG-transformed interaction for sufficiently large flow parameters $\alpha$ with a repulsive 3N contact interaction we were able to reproduce the experimental charge radius and ground-state energy systematics simultaneously. Only for cases where the two-body interaction exhibits a pathological energy systematics a contact force is clearly not sufficient to arrive at a reasonable behavior.

In a next step, we can apply these interactions in a variety of many-body schemes and study the elementary effects of 3N interactions on nuclear observables. We will employ the phenomenological 3N interactions in exact no-core shell model calculations for the spectroscopy of light nuclei and in approximate many-body approaches such as RPA for the study of the collective response of heavier nuclei. The fundamental advantage of 3N contact interactions is that the computation of three-body matrix elements itself is not a limiting factor and thus a larger range of nuclei and observables can be explored. 

A major goal, however, it to go beyond phenomenological 3N interactions for supplementing a unitarily transformed NN interaction towards a consistent two- plus three-body interaction resulting from a combined unitary transformation of an initial two- plus three-body interaction. This can be done, e.g., using an SRG-evolution of the chiral NN plus 3N interaction. A study of the ground-state energy and radius systematics for those interactions, even in a simple framework like HF+MBPT, will provide crucial information on the quality of the presently available chiral interactions for nuclear structure studies beyond the light isotopes and it will serve as a test-case for the SRG-evolution in the NN plus 3N sector. Based on our developments for phenomenological 3N interactions discussed here, we will perform a similar analysis with fully realistic NN plus 3N interactions next.

\section*{Acknowledgments}

This work is supported by the Deutsche Forschungsgemeinschaft through contract SFB 634, by the Helmholtz International Center for FAIR (HIC for FAIR) within the framework of the LOEWE program launched by the State of Hesse, by the BMBF through contract 06DA9040I, and by the GSI Helmholtzzentrum f\"ur Schwerionenforschung. 


\end{document}